\documentstyle[procl,epsfig]{article}

\input{psfig}

\def\Journal#1#2#3#4{{#1} {\bf #2}, #3 (#4)}

\def\PLB{{\em Phys. Lett.}  B}

\def\ZPC{{\em Z. Phys.} C}

\def\be{\begin{equation}}

\def\ee{\end{equation}}

\def\bea{\begin{eqnarray}}

\def\eea{\end{eqnarray}}

\begin{document}

\title{PHOTOPRODUCTION OF JETS AT HERA}

\author{ M. KLASEN}

\address{II. Institut f\"ur Theoretische Physik, Universit\"at Hamburg,
Luruper Chaussee 149, D-22761 Hamburg, Germany}

\maketitle\abstracts{
Recent NLO calculations for the photoproduction of one and two jets at HERA
are compared to data from H1 and ZEUS. We discuss the
physics potential of these measurements for constraining the photon structure,
especially the quark density at large $x$, and study remaining hadronization
and jet definition uncertainties.}

At HERA, $ep$ scattering proceeds dominantly through the radiation
of almost real photons. In a large fraction of these
events, jets with large $E_T$ are produced which allows
the application of perturbative QCD.
In LO, one has to calculate the process $ab\rightarrow 12$, where $a$ is the
photon (direct process) or a parton in the photon (resolved process),
$b$ is the parton in the proton, and $1$
and $2$ are the two outgoing partons hadronizing into two jets.
In a LO calculation one cannot produce more than two jets, and the
implementation of a jet algorithm is impossible. In addition, one suffers
from an artificial separation of direct and resolved processes and from large
scheme and scale dependences.\\

In NLO, these drawbacks are remedied through the inclusion of virtual and
real corrections. The singularities coming from the virtual loop integration
or from the integration over soft and collinear regions of the three particle
final state are regularized dimensionally and cancelled or renormalized in the
$\overline{\mbox{MS}}$ scheme. Whereas the small cone
approximation of Aurenche et al. is restricted to one-jet cross sections, the
Lorentz invariant phase space slicing method used here is very flexible
and can also be applied to two-jet cross sections. \\

\begin{figure}[ttt]
 \begin{center}
  \psfig{figure=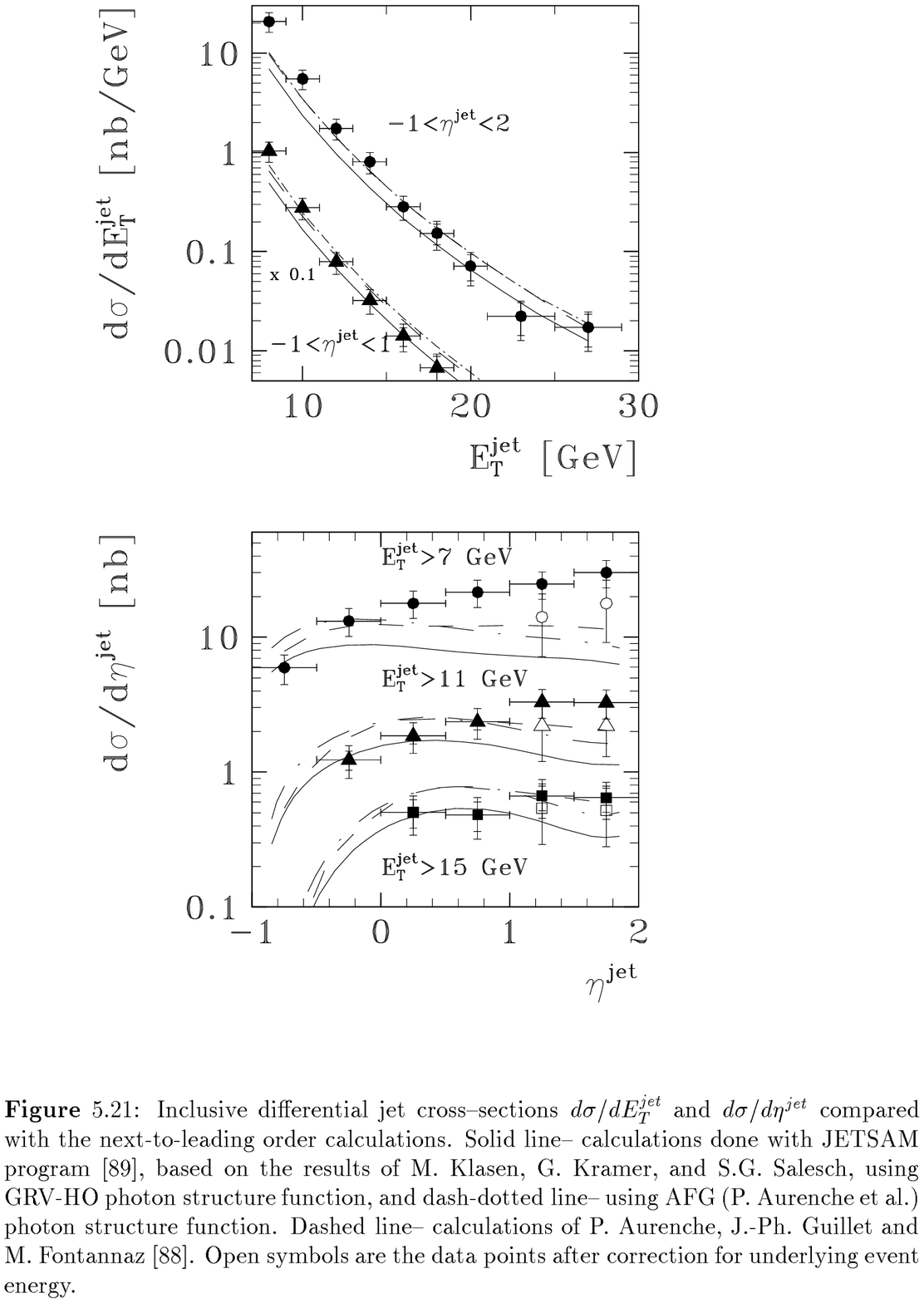,height=4cm,bbllx=133pt,bblly=491pt,bburx=415pt,%
         bbury=730pt,clip=}
  \psfig{figure=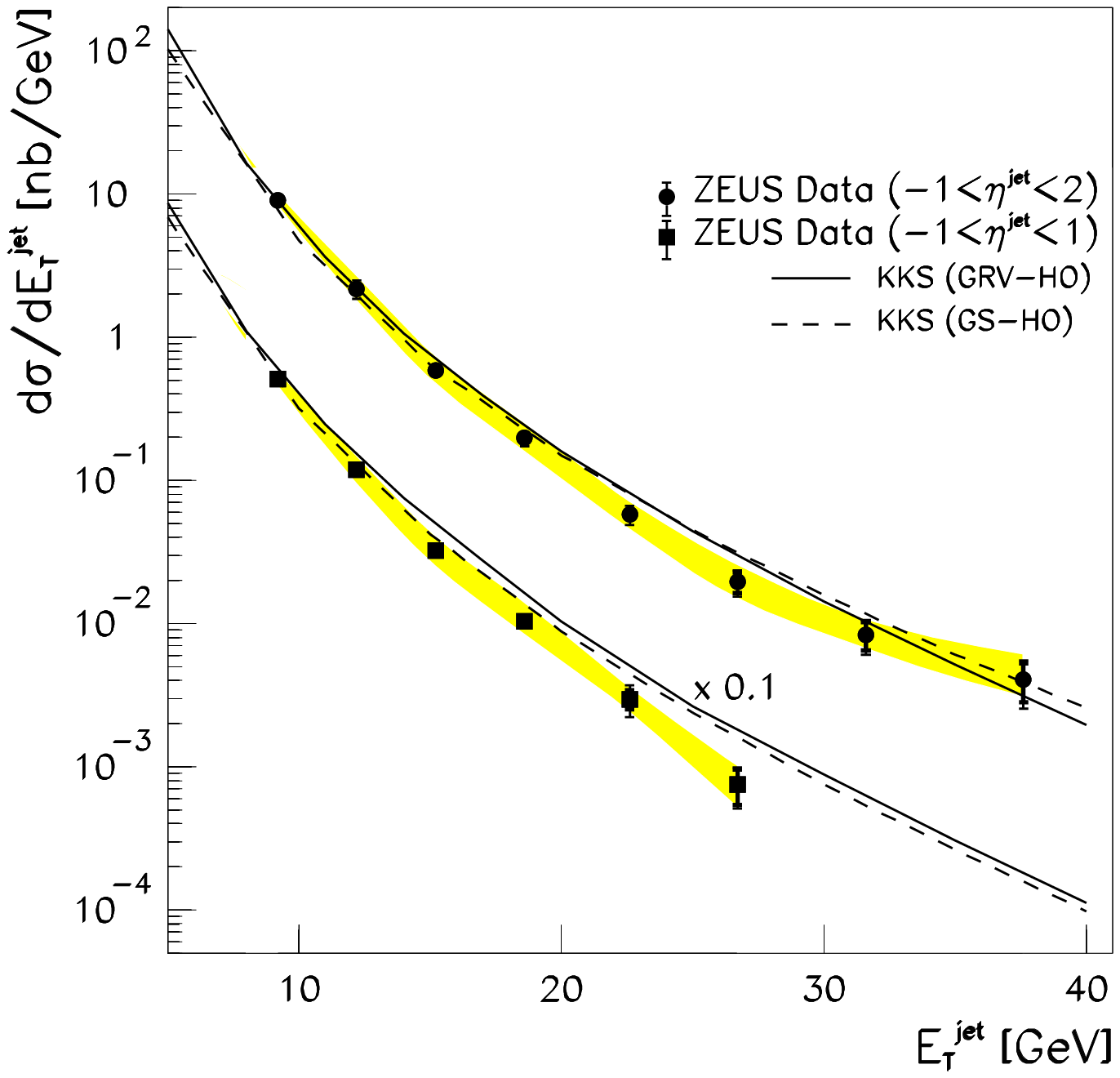,height=4cm,bbllx=72pt,bblly=210pt,bburx=466pt,%
         bbury=590pt,clip=}
  \psfig{figure=fig1ab.ps,height=4cm,bbllx=133pt,bblly=230pt,bburx=415pt,%
         bbury=468pt,clip=}
  \psfig{figure=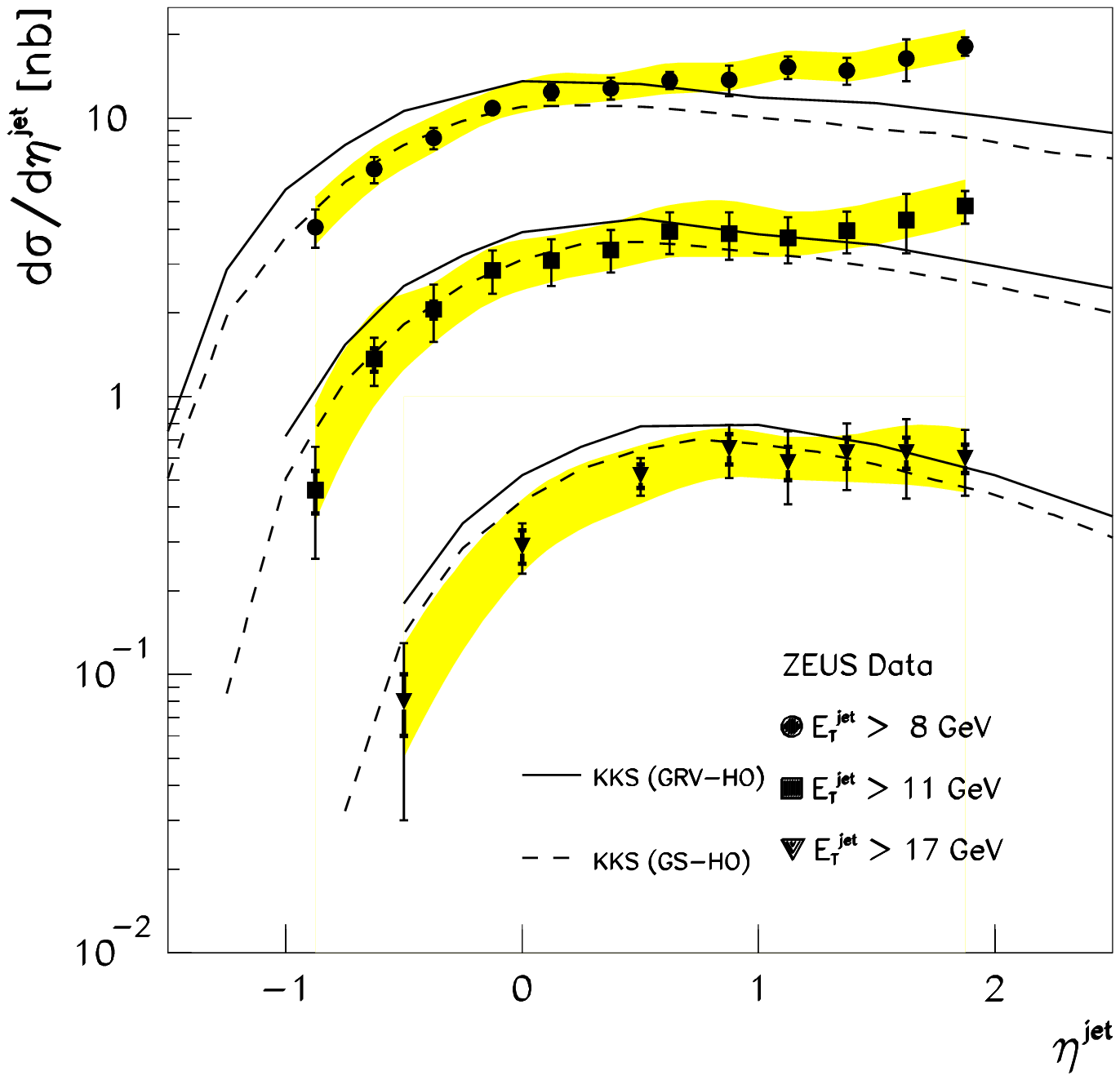,height=4cm,bbllx=72pt,bblly=210pt,bburx=466pt,%
         bbury=590pt,clip=}
  \caption{One-jet cross sections in NLO: $E_T$ (top) and $\eta$ (bottom)
           distributions compared to H1 (left) and ZEUS (right) data.}
 \end{center}
\end{figure}

\begin{figure}[ttt]
 \begin{center}
  \epsfig{figure=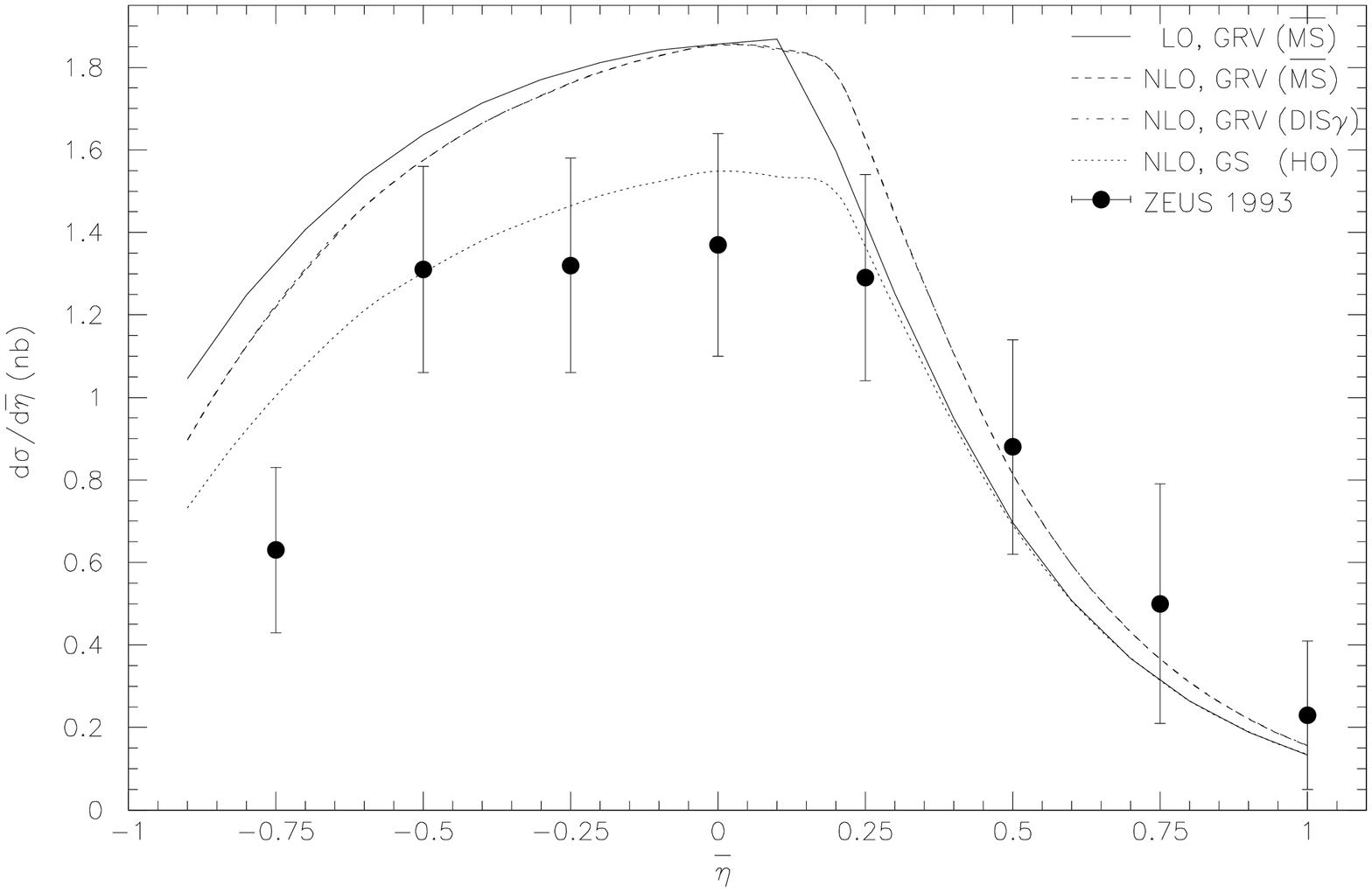,height=5.6cm,bbllx=108pt,bblly=79pt,bburx=517pt,%
         bbury=707pt,angle=-90,clip=}
  \epsfig{figure=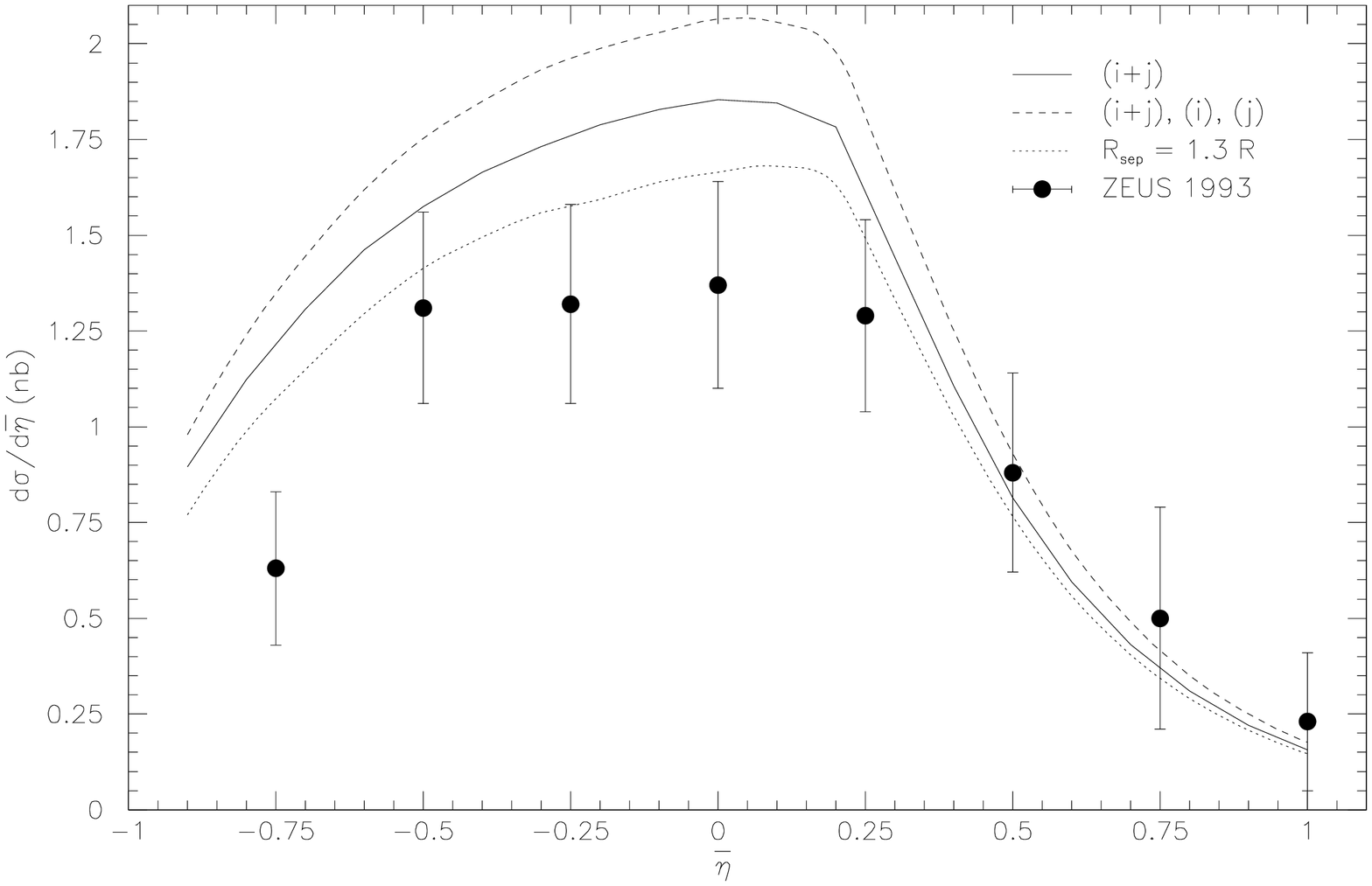,height=5.6cm,bbllx=108pt,bblly=79pt,bburx=517pt,%
         bbury=707pt,angle=-90,clip=}
  \caption{Two-jet cross sections in NLO: $\overline{\eta}$ distributions
           compared to ZEUS data for different photon structure functions
           (left) and jet definitions (right).}
 \end{center}
\end{figure}

We calculate cross sections for HERA conditions where the electron energy is
$E_e = 26.7$ GeV and the proton energy is $E_p = 820$ GeV. The spectrum
of the virtual photons is approximated by the Weizs\"acker-Williams formula
with $Q_{\max}^2 = 4~(0.01)$ GeV$^2$ and $x_a \in [0.2;0.85]~([0.25;0.7])$
for ZEUS (H1). For the proton
structure functions, we choose MRS(D-) and use the corresponding $\Lambda$
value to calculate the strong coupling $\alpha_s$ in two-loop approximation
with five flavors. Following the Snowmass convention, a jet is defined as a
bunch of hadrons contained in a cone of radius $R = 1$ in the
azimuth-rapidity plane.\\

We first compare our predictions\cite{KKS95} for one-jet cross sections to
H1 data\cite{BOU95} and predictions from Aurenche et al.\cite{AFG94}
(left hand side of fig. 1), where we
integrate either over $\eta$ (top) or $E_T$ (bottom). The agreement between
the two calculations and theory and data is excellent for the AFG structure
functions (dashed curves) except in the forward region at low $E_T$. This 
discrepancy vanishes after hadronization corrections (empty data points).
At larger $E_T$, the effect of hadronization and the discrepancy in the
forward region are reduced. The GRV photon structure function (full curves)
underestimates the H1 data. However, the ZEUS data\cite{ZEUS1}
(right hand side of fig. 1)
are overestimated when using the GRV structure function (full curves) except
in the forward region and can be better described by the GS structure function
(dashed curves). \\

Next, we turn to two-jet cross sections, where only ZEUS 1993 data\cite{ZEUS1}
and our predictions\cite{KK96a} with NLO direct and LO resolved
contributions were available so far.\footnote{For a comparison of our
predictions to ZEUS 1994 data see L. Feld's contribution to these proceedings.}
We compare the distribution in
$\overline{\eta}=\frac{1}{2}(\eta_1+\eta_2)$ integrated over $E_T > 6$ GeV and
the difference of the two rapidities $|\eta^\ast | = |\eta_1-\eta_2| < 0.5$
for $x_\gamma^{\mbox{OBS}} > 0.75$ using CTEQ(3M) structure functions in the
proton.
The ZEUS data are again overestimated by the GRV structure
function (dashed and dashed-dotted curves) and adequately described by the GS
structure function (dotted curve) (left hand side of fig. 2). Since we are at
large $x_\gamma^{\mbox{OBS}}$,
the direct process dominates with the resolved process still contributing
15\% of the total cross section, and we can distinguish mainly the quark
densities in the photon. Even with LO resolved contributions, the scheme
dependence is cancelled as can be seen when comparing the GRV predictions in
$\overline{\mbox{MS}}$ scheme (dashed curve) and in DIS$\gamma$ scheme
(dashed-dotted curve).\\

An important prerequisite for pinning down the photon structure function
is a good understanding of the jet final state. Here,
ambiguities arise due to possible double counting
of jets and restrictions in parton-parton separation. The right hand side
of fig. 2 shows these
uncertainties to be $\pm 10\%$. They are reduced at larger $E_T$ similar to
the hadronization corrections as demonstrated above for one-jet cross
sections. \\

Finally, we present the first NLO predictions for resolved photoproduction
of two jets, using again the phase space slicing method.\cite{KKK96}
The calculation was completed shortly after the workshop and checked with
our old program in one-jet production.\cite{KKS95} Fig. 3 shows the
effect of the higher order contributions for $\overline{\eta}$ distributions
in the direct dominated region (left) and the resolved dominated region
(right). As expected from one-jet production, the corrections are large
and range from 40\% (``direct'') to 75\% (``resolved'') in the central
region. Whereas they are symmetric for the latter, they
increase rapidly from the backward to the forward region for
$x_\gamma^{\mbox{OBS}} > 0.75$ thus improving the description of the data
in the forward region.

\begin{figure}[ttt]
 \begin{center}
  \epsfig{figure=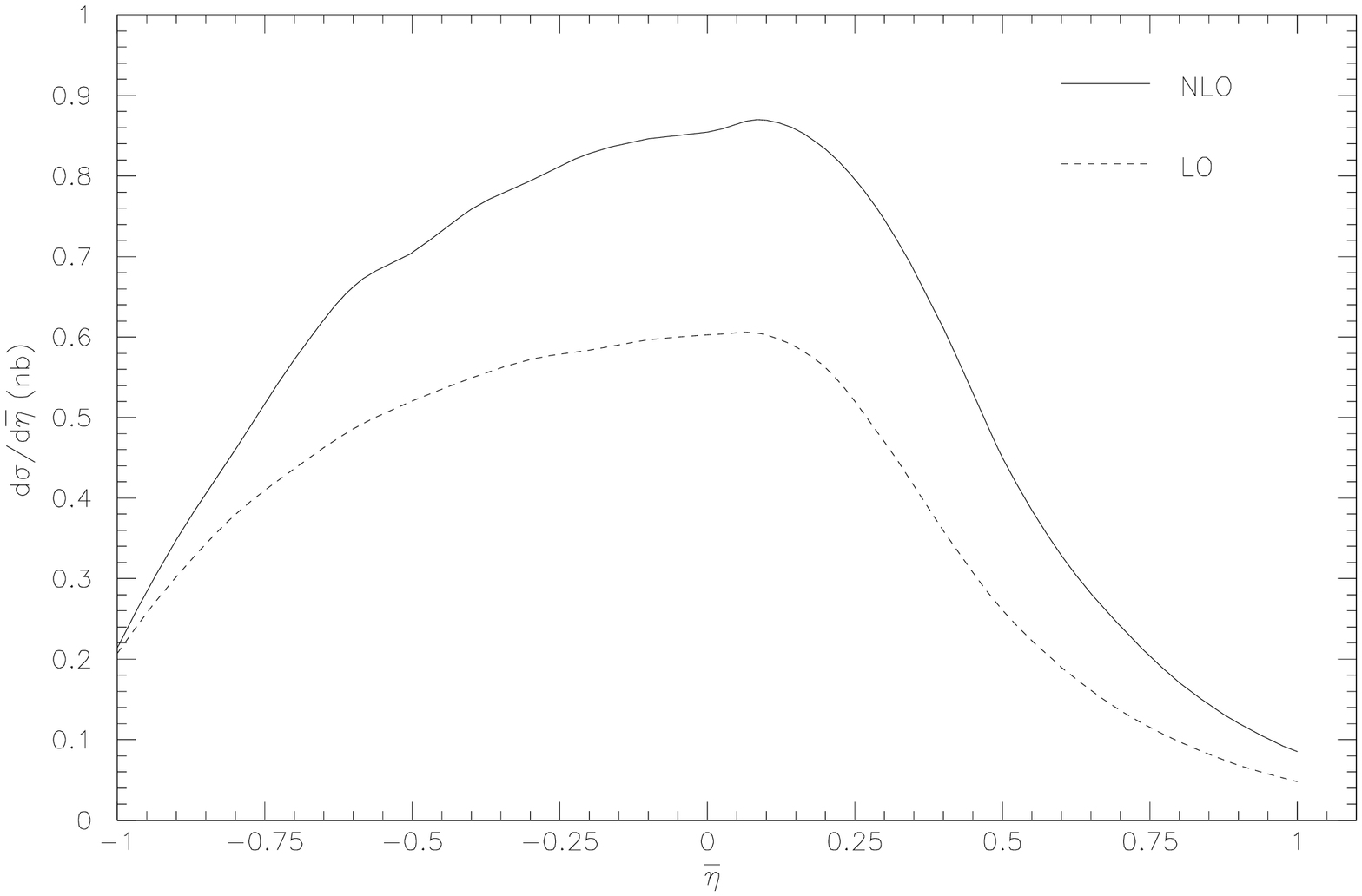,height=5.6cm,bbllx=108pt,bblly=79pt,bburx=517pt,%
         bbury=707pt,angle=-90,clip=}
  \epsfig{figure=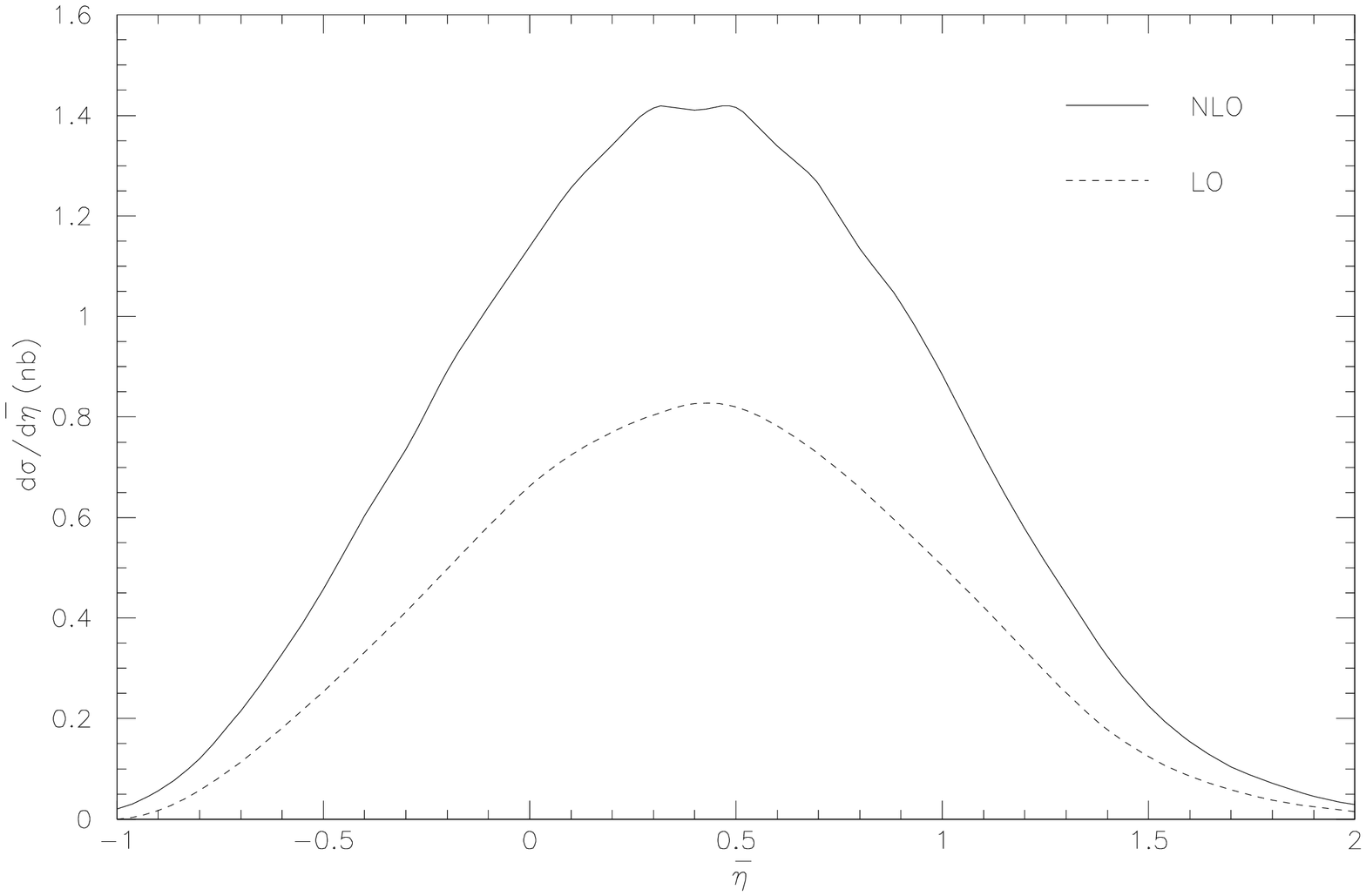,height=5.6cm,bbllx=108pt,bblly=79pt,bburx=517pt,%
         bbury=707pt,angle=-90,clip=}
  \caption{Resolved two-jet cross sections in NLO and LO: $\overline{\eta}$
           distributions for $x_\gamma^{\mbox{OBS}} > 0.75$ (left) and
           $x_\gamma^{\mbox{OBS}} \in [0.3;0.75]$ (right).}
 \end{center}
\end{figure}

\section*{Acknowledgments}
It is a pleasure to thank G. Kramer and T. Kleinwort for their collaboration
on the calculations and L. Feld and J. Butterworth for helpful discussions
concerning the jet definitions.
I would also like to thank A. Vogt and the convenors for the invitation.
This work is supported by BMBF, Bonn, Germany under contract 05\,7HH92P(0)
and EEC Program "Human Capital and Mobility" through Network "Physics at
High Energy Colliders" under Contract CHRX-CT93-0357 (DG12 COMA).


\section*{References}
\begin{thebibliography}{99}
\bibitem{KKS95} M. Klasen, G. Kramer, S.G. Salesch, \Journal{\ZPC}{68}{113}
                {1995}.
\bibitem{BOU95} A. Bouniatian, Ph.D. Thesis, DESY FH1K-95-04 (1995).
\bibitem{AFG94} P. Aurenche, J.Ph. Guillet, M. Fontannaz, \Journal{\PLB}
                {338}{98}{1994}.
\bibitem{ZEUS1} ZEUS Collaboration, \Journal{\PLB}{342}
                {417}{1995}, {\bf 348}, 665 (1995).
\bibitem{KK96a} M. Klasen, G. Kramer, \Journal{\PLB}{366}{385}{1996}.
\bibitem{KKK96} M. Klasen, T. Kleinwort, G. Kramer, to be published.
\end{thebibliography}
\end{document}